\documentclass[12pt,a4paper]{article}

 \usepackage{a4wide}
 \usepackage{latexsym}
 \usepackage{epsf}
 \usepackage{amssymb}
 \usepackage{amsmath, cite}
 \usepackage{slashed,epsfig}

\begin{document}

\begin{flushright}
\small
UG-04-01\\
{\bf hep-th/0404100}\\
\date \\
\normalsize
\end{flushright}

\begin{center}

%title

\vspace{.7cm}

{\LARGE {\bf The Domain Walls of }} \\
~ \\
{\LARGE {\bf Gauged Maximal Supergravities }} \\
~ \\
{\LARGE {\bf and their M-theory Origin }}

\vspace{1.2cm}

%authors
{\large Eric Bergshoeff, Mikkel Nielsen and Diederik Roest} \vskip 1truecm

\small
%institutes
{\it Centre for Theoretical Physics, University of Groningen,\\
   Nijenborgh 4, 9747 AG Groningen, The Netherlands\\
E-mail: {\tt (e.a.bergshoeff, m.nielsen, d.roest)@phys.rug.nl}}

\vspace{.7cm}

%%%%%%%%%%%%%%%%%%%%%%%%%%%%%%%%%%%%%%%%%%%%%%%%%%%%%%%%%%%%%%%%%%%%%%

{\bf Abstract}

\end{center}

\begin{quotation}

\small

We consider gauged maximal supergravities with $CSO(p,q,r)$ gauge
groups and their relation to the branes of string and M-theory.
The gauge groups are characterised by $n$ mass parameters, where
$n$ is the transverse dimension of the brane. We give the scalar
potentials and construct the corresponding domain wall solutions.
In addition, we show the higher-dimensional origin of the domain
walls in terms of (distributions of) branes.

We put particular emphasis on the $CSO(p,q,r)$ gauged
supergravities in $D=9$ and $D=8$, which are related to the
D7-brane and D6-brane, respectively. In these cases, twisted and
group manifold reductions are shown to play a crucial role. We
also discuss salient features of the corresponding brane
distributions.

\end{quotation}

\newpage

\pagestyle{plain}

\section{Introduction}

Due to the AdS/CFT correspondence \cite{Maldacena:1998re}, it has
been realised that there is an intimate relationship between
certain branes of string and M-theory and lower-dimensional
$SO(n)$ gauged supergravities. The relation is established via a
maximally supersymmetric vacuum configuration of string or
M-theory, which is the direct product of an AdS space and a
sphere. For a $p$-brane with $n$ transverse directions we are
dealing with an $AdS_{p+2} \times S^{n-1}$ vacuum configuration,
where $p+n+1 = 10$ or $11$. On the one hand, this vacuum
configuration arises as the near-horizon limit of the $p$-brane in
question; on the other hand, the coset reduction over the sphere
part leads to the related $SO(n)$ gauged supergravity in $p+2$
dimensions which allows for a maximally supersymmetric $AdS_{p+2}$
vacuum configuration. The gauge theory of the AdS/CFT
correspondence can be taken at the boundary of this $AdS_{p+2}$
space. All lower-dimensional dilatons are vanishing for this
vacuum configuration. This is related to the conformal invariance
of the gauge theory. For the branes occurring in the AdS/CFT
correspondence, the situation is summarized in the table below.

\begin{table}[ht]
\begin{center}
\hspace{-1cm}
\begin{tabular}{||c|c|c||}
\hline \rule[-1mm]{0mm}{6mm}
  Brane & Vacuum configuration & Gauged SUGRA \\
\hline \hline \rule[-1mm]{0mm}{6mm}
 M2&$AdS_4 \times S^7$&$D=4$ $SO(8)$\\
\hline \rule[-1mm]{0mm}{6mm}
 M5&$AdS_7 \times S^4$&$D=7$ $SO(5)$\\
\hline \rule[-1mm]{0mm}{6mm}
 D3&$AdS_5 \times S^5$&$D=5$ $SO(6)$\\
\hline
\end{tabular}
\caption{\it The branes of the AdS/CFT correspondence.}
\end{center}
\end{table}

There are two ways to depart from the conformal invariance, which both involve exciting
some of the dilatons in the vacuum configuration. For the cases given in the table, the
scalar potential of the gauged supergravity contains $n-1$ dilatons. By exciting some of
these dilatons one obtains a deformed AdS vacuum configuration, which can also be seen as
a domain wall. In the AdS/CFT correspondence this corresponds to considering the
(non-conformal) Coulomb branch of the gauge theory, see e.g.~\cite{Kraus:1998hv,
Freedman:1999gk, Cvetic:1999xx}. Alternatively, one can obtain a non-conformal theory by
considering the other branes of string and M-theory, for which there is an extra dilaton
present in the scalar potential of the gauged supergravities. In these cases the
(maximally supersymmetric) AdS vacuum is replaced by a (non-conformal and
half-supersymmetric) domain wall solution. This situation is encountered when one tries
to generalise the AdS/CFT correspondence to a DW/QFT correspondence
\cite{Itzhaki:1998dd,Boonstra:1998mp,Behrndt:1999mk}.

A natural generalisation is to excite some of the $n-1$ dilatons,
leading to the Coulomb branch of the CFT, and the extra dilaton
that leads to a non-conformal QFT {\sl simultaneously}. This gives
rise to domain wall solutions of $SO(n)$ gauged supergravities
\cite{Cvetic:2000zu} that describe the Coulomb branch of the
(non-conformal) QFT. The uplift of these domain walls leads to the
near-horizon limit of brane distributions in string and M-theory.

It is the purpose of this paper to extend the analysis of \cite{Cvetic:2000zu} to the
case of $CSO(p,q,r)$ gauged supergravity theories. In doing this we will see that an
interesting pattern emerges where all $CSO(p,q,r)$ gauged supergravities can be treated
in a unified way. We will pay particular attention to the branes with a small number of
transverse directions, i.e.~with $n \le 3$. As we will see for these cases, we will
encounter not only coset reductions on a sphere but also group manifold reductions on
$S^3$ and twisted $SL(2,\mathbb{R})$ reductions on $S^1$.

The organisation of the paper is as follows. In section \ref{CSO-gaugings} we briefly
review some salient features of $CSO(p,q,r)$ gauged supergravities and their construction
by dimensional reduction. The scalar potentials of these theories are separately
discussed in section \ref{scalar potentials}. The domain wall solutions of $CSO(p,q,r)$
gauged supergravities are presented in section \ref{mdw}. The higher-dimensional origin
of these solutions as brane solutions in terms of a harmonic  over the (flat) transverse
space $\mathbb{R}^n$ is discussed in section \ref{hd}. In section \ref{bd} we restrict to
the case where the gauge group is $SO(n)$ or a contracted version thereof and give the
higher-dimensional interpretation as brane distributions. We discuss our results in
section \ref{discussion}. Finally, in appendix \ref{contraction} we perform two limiting
procedures on the scalar sector of $CSO(p,q,r)$ gauged supergravity theories, which are
used in the main text.

\section{$CSO(p,q,r)$ Gauged Supergravities} \label{CSO-gaugings}

It has been known for long that certain gauged maximal supergravities with global
symmetry groups $SL(n,\mathbb{R})$ allow for the gauging of the $SO(n)$ subgroup of the
global symmetry. An example is the $SO(8)$ gauging in four dimensions
\cite{deWit:1982ig}. Subsequently, it was realised that such gauged supergravities could
be obtained by the reduction of a higher-dimensional supergravity over a sphere, with a
flux of some field strength through the sphere. For example, the $SO(8)$ theory can be
constructed by the reduction of 11D supergravity over $S^7$, with magnetic flux of the
four-form field strength through the seven-sphere \cite{deWit:1987iy}. Other examples are
given in table\footnote{The $S^5$ reduction of IIB has not (yet) been proven in full
generality. The linearised result was obtained in \cite{Kim:1985ez} while the full
reduction of the $SL(2,\mathbb{R})$ invariant part of IIB supergravity was performed in
\cite{Cvetic:2000nc}.}~\ref{tab:gaugings}.

\begin{table}[ht]
\begin{center}
\hspace{-1cm}
\begin{tabular}{||c|c|c|c|c||}
\hline \rule[-1mm]{0mm}{6mm}
  $D$ & $n$ & $\phi$ & Origin & Brane\\
\hline \hline \rule[-1mm]{0mm}{6mm}
  $10$ & $1$ & $\surd$ & Massive IIA \cite{Romans:1986tz}&D8\\
\hline \rule[-1mm]{0mm}{6mm}
  $9$ & $2$ & $\surd$ & IIB with $SO(2)$ twist \cite{Meessen:1998qm}&D7 \\
\hline \rule[-1mm]{0mm}{6mm}
  $8$ & $3$ & $\surd$ & IIA on $S^2$ \cite{Salam:1985ft}&D6 \\
\hline \rule[-1mm]{0mm}{6mm}
  $7$ & $5$ & $-$ & 11D on $S^4$ \cite{Nastase:1999cb, Nastase:1999kf}&M5 \\
\hline \rule[-1mm]{0mm}{6mm}
  $6$ & $5$ & $\surd$ & IIA on $S^4$ \cite{Cvetic:2000ah}&D4 \\
\hline \rule[-1mm]{0mm}{6mm}
  $5$ & $6$ & $-$ & IIB on $S^5$ \cite{Kim:1985ez, Cvetic:2000nc}&D3 \\
\hline \rule[-1mm]{0mm}{6mm}
  $4$ & $8$ & $-$ & 11D on $S^7$ \cite{deWit:1987iy}&M2 \\
\hline \rule[-1mm]{0mm}{6mm}
  $3$ & $8$ & $\surd$ & IIA on $S^7$ \cite{Fischbacher:2003yw} &F1A\\
\hline  \rule[-1mm]{0mm}{6mm}
  $2$ & $9$ & $\surd$ & IIA on $S^8$ \cite{Nicolai:2000zt} &D0\\
\hline \hline \rule[-1mm]{0mm}{6mm}
  $1$ & $10$ & $\surd$ & IIB$_\text{E}$ on $S^9$ &D-instanton\\
\hline
\end{tabular}
\caption{\it This table indicates the $D$-dimensional gauged maximal supergravities and
the corresponding $(D-2)$-branes, with $n$ the number of mass parameters as well as the
number of transverse directions. The third column indicates whether the scalar potential
depends on the extra dilaton $\phi$; the fourth column gives the higher-dimensional
origin of the $SO(n)$ prime examples. The last row corresponds to a (conjectured)
reduction of Euclidean IIB supergravity on a nine-sphere. \label{tab:gaugings}}
\end{center}
\end{table}

In addition to $SO(n)$, the global symmetry group $SL(n,\mathbb{R})$ has more subgroups
that can be gauged. It was found that many more gaugings could be obtained from the
$SO(n)$ prime examples by analytic continuation or group contraction of the gauge group
\cite{Hull:1984vg, Hull:1984qz}. This leads one from $SO(n)$ to the group\footnote{The
generalisation of $SO(p,q) \equiv CSO(p,q,0)$ to $CSO(p,q,r)$ is non-trivial for
odd-dimensional gauged supergravities; a crucial role is played
\cite{Andrianopoli:2000fi} by the massive self-dual gauge potentials in odd dimensions
\cite{Townsend:1984xs}.} $CSO(p,q,r)$ with $p+q+r = n$. These gaugings are described in
terms of a symmetric matrix $Q$, which can always be diagonalised as $Q={\rm
diag}(q_1,\ldots,q_n)$. Furthermore, the individual $q_i$'s can always be chosen to be
$\pm 1$ or $0$:
 \begin{align}
  Q = \left( \begin{array}{ccc}
                       \mathbb{I}_p & 0 & 0 \\
                       0 & -\mathbb{I}_q & 0 \\
                       0 & 0 & 0_r
                      \end{array} \right) \,.
  \label{general-Q}
 \end{align}
This generalises the $SO(n)$ gauging to $[ n^2 / 4 + n ]$ different possible gaugings.

%For example, the resulting field content in $D=5$ contains $15+r$
%gauge vectors and $12-r$ massive self-dual two-form potentials
%\cite{Andrianopoli:2000fi}. In $D=7$ one would expect $r$ massless
%two-forms and $5-r$ massive self-dual three-forms, of which the
%case $r=1$ is confirmed in \cite{Cvetic:2000ah}. Surprisingly,
%this phenomenon does not occur in $D=9$, where one has one
%massless three-form potential for all values of $r$
%\cite{Meessen:1998qm}. This is related to the fact that the 9D
%potential is a singlet, while the lower-dimensional potentials
%transform non-trivially under the gauge group. In this paper, we
%will be concerned with the scalar subsector of these theories and
%therefore not mind the subtleties associated with the gauge
%potentials.

The question of the higher-dimensional origin of the $CSO$
gaugings was clarified in \cite{Hull:1988jw}, where the same
operations of analytic continuations and group contractions were
applied to the internal manifold. The resulting manifolds are
hypersurfaces in $\mathbb{R}^n$ (with Cartesian coordinates
$\mu_i$) defined by
 \begin{align}
  \sum_{i=1}^n q_i \mu_i{}^2 = 1 \,,
 \label{hypersurface}
 \end{align}
where $q_i = 0 , \pm 1$ are the diagonal entries of the matrix $Q$
\eqref{general-Q}. The manifold corresponding to
\eqref{hypersurface} is denoted by \cite{Hull:1988jw}
\begin{align}
H^{p,q} \circ T^r\,,
\end{align}
where we use the symbol $\circ$ instead of $\times$ to indicate
that the manifold is not a direct product. The corresponding
reduction should first be performed over the toroidal part,
followed by the hyperbolic manifold $H^{p,q}$. The latter can be
endowed with a positive-definite metric, which generically is
inhomogeneous \cite{Cvetic:2004km}; the exceptions are the
(maximally symmetric) coset spaces
 \begin{align}
  S^{n-1} = H^{n,0} \simeq \frac{SO(n)}{SO(n-1)} \,, \qquad
  H^{n-1} = H^{1,n-1} \simeq \frac{SO(1,n-1)}{SO(n-1)} \,,
 \end{align}
i.e.~the sphere and the hyperboloid. Generically the spaces $H^{p,q}$ are non-compact;
the only exception is the sphere with $q=0$.

Thus non-compact gauge groups $CSO(p,q,r)$ with $q \neq 0$ are obtained from reduction
over non-compact manifolds, as first suggested in \cite{Pernici:1985nw}. It can be argued
that the corresponding reduction, albeit a non-compactification, is a consistent
truncation provided the compact case with $q = 0$ has been proven consistent
\cite{Hull:1988jw}.

We now discuss the special case of $p+q \leq 2 $. The manifold
\eqref{hypersurface} is then understood as a one-dimensional
manifold $S^1$. It is instructive to consider a chain of
contractions, starting from a coset reduction over an
$(n-1)$-sphere yielding an $SO(n)$ gauge group\footnote{Note that
the manifold is unchanged in the last contraction.}:
 \begin{align}
   S^{n-1} \rightarrow S^{n-2} \circ S^1 \rightarrow \cdots
   \rightarrow S^2\circ T^{n-3}\rightarrow
   S^1 \circ T^{n-2} \rightarrow S^1 \circ T^{n-2} \,.
 \label{chaina}
 \end{align}
From \eqref{general-Q} and \eqref{hypersurface}, the resulting gauge groups should be
 \begin{align}
   (n,0,0) \rightarrow (n-1,0,1) \rightarrow \cdots \rightarrow (3,0,n-3)\rightarrow (2,0,n-2) \rightarrow
   (1,0,n-1)\,,
 \label{chainb}
 \end{align}
where we have used a short-hand notation $(p,0,r)$ for the group
$CSO(p,0,r)$. The following puzzle arises for the last two links
of this chain: the reduction over the torus $S^1 \circ T^{n-2}$ is
supposed to lead to the gauge group $CSO(2,0,n-2)$ and its further
contraction $CSO(1,0,n-1)$, while toroidal reduction na\"{i}vely
leads to an ungauged theory with trivial gauge group $U(1)^{n-1}$.

This puzzle is yet more apparent when we consider the simplest case of $n=2$. In this
case the reduction of IIB over $S^1$ is supposed to lead to the gauge group $SO(2)$ or
its contraction $\mathbb{R}$. The resolution lies in a {\sl twisted} reduction over $S^1$
\cite{Scherk:1979ta}, making use of the $SL(2,\mathbb{R})$ duality group of IIB
supergravity (see \cite{Bergshoeff:1996ui, Lavrinenko:1998qa, Meessen:1998qm,
Hull:2002wg, Bergshoeff:2002mb, Cowdall:2000sq}). These twisted reductions give rise to
$CSO$ gauged supergravity in 9D with $n=2$. The three different possibilities correspond
to twisting with the subgroup $SO(2)$ of $SL(2,\mathbb{R})$, the analytic continuation
$SO(1,1)$ and the contraction $\mathbb{R}$, respectively\footnote{In the case of $SO(2)$
twisting, the reduction can be given an alternative interpretation as a Kaluza-Klein
reduction with a consistent truncation to a higher mode \cite{Dabholkar:2002sy}.}.

The reduction Ansatz for a twisted reduction involves an $SL(2,\mathbb{R})$
transformation of the form \cite{Meessen:1998qm}
 \begin{align}
  \Omega = \exp(C y) \,,
 \end{align}
where $y$ is the internal coordinate and $C$ is a traceless matrix. Note that general
$SL(n,\mathbb{R})$ twisted reductions \cite{Scherk:1979ta, Dabholkar:2002sy} give rise to
a traceless matrix $C$. Only when twisting with an $SL(2,\mathbb{R})$ subgroup of
$SL(n,\mathbb{R})$ can one relate the traceless matrix via
 \begin{align}
  C_p{}^q = \epsilon_{pr} Q^{qr} \,, \qquad
  Q^{qr} = \text{diag} (q_1,q_2) \,,
  \label{traceless}
 \end{align}
to a symmetric matrix $Q$. Due to \eqref{traceless}, the traceless
matrix $C$ can be related to the para-meters %%% NOTE "-" inserted to avoid param-eters
$q_1$ and $q_2$ of \eqref{hypersurface}. The explicit relation
between $y$ and the Cartesian coordinates $\mu_i$ reads
 \begin{align}
  \mu_1 = \text{sin}(\sqrt{q_1 q_2} y) / \sqrt{q_1} \,, \qquad
  \mu_2 = \text{cos}(\sqrt{q_1 q_2} y) / \sqrt{q_2} \,.
 \end{align}
This explains the relation between twisted reduction and the case $p+q \leq 2$ of
\eqref{hypersurface}.

The cases $n>2$ can be treated in a similar way: at the two last links of the above chain
of contractions, one must perform an $SL(2,\mathbb{R})$-twisted reduction. The required
$SL(2,\mathbb{R})$-symmetry is always present for $n \ge 4$, because any reduction over a
$T^2$ factor yields this symmetry. For instance, for $n=4$ we have:
\begin{align}
 S^3 \rightarrow S^{2} \circ S^1 \rightarrow S^1 \circ {T^2}
 \,.
\end{align}
The reduction over the two-torus gives rise to the $SL(2,\mathbb{R})$-symmetry that is
needed to perform the twisted reduction over the remaining $S^1$. The same happens for
all cases $n \ge 4$. The difference between $(p,q,r) = (2,0,n-2)$, $(1,1,n-2)$ and
$(1,0,n-1)$ is the flux of the scalars: the different values correspond to twisting with
the subgroups $SO(2)$, $SO(1,1)$ and $\mathbb{R}$ of $SL(2,\mathbb{R})$, respectively.

The $n=3$ case needs special attention. In this case we are
dealing with two-dimensional spaces, e.g.~$S^2$ and $H^2$, over
which one can perform coset reductions. However, by contraction we
get
\begin{align} \label{chain2}
  S^2\ {\rm or}\ H^2\ \rightarrow\ S^1 \circ S^1\,,
\end{align}
and we seemingly lack the $SL(2,\mathbb{R})$-symmetry due to the
absence of a $T^2$ factor. However, the case $n=2$ has a peculiar
feature: the reduction over $S^2$ or $H^2$ (i.e.~the first link of
\eqref{chain2}) is only allowed for theories that have an origin
in one dimension higher. From the higher-dimensional point of
view, the reduction Ansatz over $S^2$ or $H^2$ corresponds to a
group manifold reduction \cite{Boonstra:1998mp, Cvetic:2003jy}:
\begin{align} \label{chain3}
  \mathcal{G}^{\text{3D}} = {(S^2\ {\rm or}\ H^2) \circ S^1} \rightarrow S^1 \circ T^2
  \,.
\end{align}
Due to the hidden higher-dimensional origin, a two-torus appears
on the right hand side, allowing for an $SL(2,\mathbb{R})$ twisted
reduction over the remaining circle.

The higher-dimensional connection corresponds to the relation
between M-theory and type IIA. One can either perform a
two-dimensional coset reduction of IIA or a three-dimensional
group manifold reduction of 11D to obtain the $SO(3)$ and
$SO(2,1)$ gauged supergravities in 8D \cite{Salam:1985ft,
AlonsoAlberca:2003jq}. The contracted versions are obtained by
first reducing 11D over $T^2$ to 9D which produces an
$SL(2,\mathbb{R})$ symmetry in 9D. Next, one applies a twisted
reduction form 9D to 8D. Twisting with the subgroups $SO(2),
SO(1,1)$ and $\mathbb{R}$ of $SL(2,\mathbb{R})$ leads to $D=8$
gauged supergravities with gauge groups $ISO(2), ISO(1,1)$ and
Heisenberg, respectively.

In the reduction \eqref{chain3} one uses group manifolds of class A of the Bianchi
classification, whose structure constants can be written as
 \begin{align}
  f_{mn}{}^p = \epsilon_{mnq} Q^{pq} \,, \qquad
  Q^{mn} = \text{diag} (q_1,q_2,q_3) \,.
 \label{structure-constants}
 \end{align}
Note that the structure constants can only be written in terms of a symmetric matrix $Q$
for {\sl three-dimensional} group manifolds. In the group manifold reduction, the
internal metric is described in terms of the Maurer-Cartan 1-forms
$\sigma^m=U^m{}_n\,dy^n$, which in turn combine into the structure constants
\eqref{structure-constants} of the 8D gauged supergravity and therefore the mass
parameters $q_i$. For more details, see~\cite{AlonsoAlberca:2003jq}.

Reducing from 11D to 10D, one finds a relation between the
three-dimensional group manifold  reductions  (with coordinates
$y^1, y^2, y^3$) and the reductions over the two-dimensional
hypersurface \eqref{hypersurface}, which boils down to the
following expression for the Cartesian coordinates
\begin{align}\nonumber
\mu_1&=\sin(\sqrt{q_2 q_3}\,y^2)/\sqrt{q_1}\,,\\
\mu_2&=\sin(\sqrt{q_1 q_3}\,y^1)\cos(\sqrt{q_2
q_3}\,y^2)/\sqrt{q_2}\,,\\\nonumber \mu_3&=\cos(\sqrt{q_1
q_3}\,y^1)\cos(\sqrt{q_2 q_3}\,y^2)/\sqrt{q_3}\,,
\end{align}
where $y^{1,2}$ are the two coordinates of the 3D group manifold
that remain after reduction over $y^3$ to 10D.

\section{Scalar Potentials} \label{scalar potentials}

We consider truncations of maximal gauged supergravities to the
sector with only gravity and the dilatons. The consistency of such
truncations have been discussed in
e.g.~\cite{Cvetic:1999xx,Cvetic:2000zu}. The corresponding
Lagrangian in $D$ dimensions is given by the kinetic terms and a
scalar potential (due to the gauging):
 \begin{align}
  & {\mathcal L} = \sqrt{-g} ( R -\tfrac{1}{2} (\partial \phi)^2 + \tfrac{1}{4}
  \text{Tr}[\partial M \partial M^{-1}] - V) \,, \qquad
  M = \text{diag} (e^{\vec{\alpha}_1 \cdot \vec{\phi}},
\ldots,e^{\vec{\alpha}_n \cdot
  \vec{\phi}}) \,.
 \label{Lagrangian}
 \end{align}
The scalars in $M$ are a truncated parametrisation of the scalar coset of the particular
maximal supergravity we are considering. In all cases this scalar coset $G/H$ will be of
the form
 \begin{align}
  M \in \frac{SL(n,\mathbb{R})}{SO(n)} \,,
 \end{align}
and it is described by the $n$ vectors $\vec{\alpha}_i=\{\alpha_{iI}\}$, which are
weights of $SL(n,\mathbb{R})$ and satisfy the following relations
 \begin{align}
  \sum_{i=1}^n \alpha_{iI}=0\,,\quad \sum_{i=1}^n
  \alpha_{iI}\,\alpha_{iJ}=2\,\delta_{IJ}\,,\quad
  \vec{\alpha}_{i}\cdot\vec{\alpha}_{j}=2\,\delta_{ij}-\frac{2}{n} \,, \label{weights}
 \end{align}
with indices $i,j=(1,\ldots,n)$ and $I,J = (1,\ldots,n-1)$. In addition we allow for an
extra dilaton $\phi$, which would correspond to an extra $\mathbb{R}^+$ factor on the
scalar manifold. It is present in some maximal supergravities and absent in others.
Explicit information on $n$, $D$ and $\phi$ can be found in table \ref{tab:gaugings}.

Note that $M$ and $\phi$ generically do not describe the full
scalar coset of maximal supergravities, however, they do
constitute the part that is relevant to the $CSO$ gauging and
scalar potential. Similarly, the full global symmetry will often
be larger than $SL(n,\mathbb{R})$. Its $SL(n,\mathbb{R})$ subgroup
will generically be the largest symmetry of the Lagrangian,
however, and is the only part of the symmetry group that is
relevant for the present discussion.

The corresponding scalar potential, coming from the gauging of the
group $CSO(p,q,r)$, takes the following form
 \begin{align}
  V = e^{a \phi} \Big(\text{Tr}[QMQM]- \tfrac{1}{2} (\text{Tr}[QM])^2\Big) \,, \qquad
  Q = \text{diag} (q_1,\ldots,q_n) \,,
  \label{CSO-potential}
 \end{align}
in terms of $n$ mass parameters $q_i$. The dilaton coupling $a$ is given by
 \begin{align}
  a^2 = \frac{8}{n} - 2 \, \frac{D-3}{D-2} \,, \label{Dna}
 \end{align}
for the different cases. For later purposes, it is convenient to
write the potential $V$ in terms of the superpotential $W$
\begin{align}
 V = \tfrac{1}{2} (\vec{\partial} W)^2 + \tfrac{1}{2} (\partial_\phi W)^2 - \frac{D-1}{4(D-2)} \, W^2 \,, \qquad
 W = e^{a\phi/2}\,\text{Tr}[QM] \,,
 \label{CSO-superpotential}
\end{align}
where $\vec{\partial} W=\partial W/\partial \vec{\phi}$ and $\partial_\phi W=\partial
W/\partial \phi$. Note that this form is only possible for dilaton couplings that satisfy
\eqref{Dna}.

In accordance with table~\ref{tab:gaugings}, $a$ vanishes for
$(D,n) = (7,5)$, $(5,6)$ and $(4,8)$, for which the extra dilaton
$\phi$ is absent. The $SL(2,\mathbb{R})$ twisted reduction of IIB
\cite{Meessen:1998qm, Bergshoeff:2002mb} and class A group
manifold reduction of 11D \cite{AlonsoAlberca:2003jq} yield scalar
potentials that coincide with \eqref{CSO-potential} for $(D,n) =
(9,2)$ and $(8,3)$, respectively. In addition, the scalar
potential of massive IIA supergravity \cite{Romans:1986tz} is also
of exactly this form with $(D,n)=(10,1)$ and is therefore included
in table~\ref{tab:gaugings}.

For the $SO(n)$ cases, i.e.~all $q_i = 1$, the scalar subsector
can be truncated by setting $M = \mathbb{I}$. In this truncation,
the scalar potential reduces to a single exponential potential
 \begin{align}\label{V-single}
  V = - \tfrac{1}{2} n (n-2) e^{a \phi} \,.
 \end{align}
Note the dependence of the sign of the potential on $n$: it is
positive for $n=1$, vanishing for $n=2$ and negative for $n \geq
3$. If $a=0$ (which necessarily implies $n \geq 3$ in $D \geq 4$),
the scalar potential becomes a cosmological constant and allows
for a fully supersymmetric AdS solution; for this reason, such
theories are called AdS supergravities. Theories with $a \neq 0$
are called DW supergravities since the natural vacuum is a domain
wall solution.

In the appendix the dimensional reduction and group contraction (which amounts to setting
one $q_i$ to zero) of the scalar potential are discussed. We show that the only effect of
these operations is to decrease $D$ or $n$ by one, respectively: the resulting system
still satisfies all equations, including \eqref{Dna} for the dilaton coupling $a$, with
the new values of the parameters $D$, $n$ and $a$. This proves that the scalar subsectors
of different gauged supergravities reduce onto each other upon performing dimensional
reductions and/or group contractions. We expect this to hold for the full theories as
well.

This expectation is supported by the following facts seen from the brane point of view. A
$p$-brane can be reduced in two ways: via a double dimensional reduction (leading to a
($p-1$)-brane in one dimension lower) or a direct dimensional reduction (leading to a
$p$-brane in one dimension lower). It has been pointed out \cite{Boonstra:1998mp} that
direct dimensional reduction leads from $SO(n)$ gauged supergravities to the contractions
thereof, which fit into the $CSO$ gauged supergravities of \cite{Hull:1984vg,
Hull:1984qz}. Thus, we get the following relations between operations on the brane and
the gauged supergravity:
\begin{center}
\begin{tabular}{ccc}
${\underline {\rm Brane}}$ && ${\underline {\rm Gauged\ supergravity}}$\\
%&& \\
\rule[-1mm]{0mm}{8mm}
direct\ dimensional\ reduction & $\Leftrightarrow$ & group contraction\\
double\ dimensional\ reduction & $\Leftrightarrow$ & circle
reduction
\end{tabular}
\end{center}

\section{Domain Walls} \label{mdw}

In this section, we give a unified description for a class of domain wall solutions for
gauged supergravities in various dimensions. We consider the following Ansatz for the
domain wall with $D-1$ world-volume coordinates $\vec{x}$ and one transverse coordinate
$y$:
 \begin{align}
   ds^2 & = g(y)^2 d\vec{x}^2 + f(y)^2 dy^2 \,, \qquad
   M = M(y) \,, \qquad \phi = \phi (y) \,.
 \label{Ansatz}
 \end{align}
The idea is to substitute this Ansatz into the action and write it
as a sum of squares, as was done for the conformal cases, i.e.~all
$q_i=1$ and $a=0$, in \cite{Bakas:1999fa}. Using
\eqref{CSO-superpotential}, the reduced one-dimensional action can
be written as
\begin{align}\nonumber
  S = \int dy \,
    g^{D-1} f \Big[ & \frac{D-1}{4(D-2)}\Big(\frac{2(D-2)}{fg}\frac{dg}{dy}- W\Big)^2
    -\frac{1}{2}\Big(\frac{1}{f}\frac{d\vec{\phi}}{dy}+\vec{\partial} W\Big)^2+\\
& -\frac{1}{2} \Big(\frac{1}{f}\frac{d\phi}{dy}+\partial_\phi W\Big)^2
+\frac{1}{f}\,\frac{dW}{dy}+(D-1)\frac{1}{fg}\frac{dg}{dy}W\Big]\,, \label{Bogomolnyi}
\end{align}
which is a sum of squares, up to a boundary term. Minimalisation of this action gives
rise to the first-order Bogomol'nyi equations
\begin{align}
  \frac{1}{f}\frac{d\vec{\phi}}{dy}=-\vec{\partial}W\,,\quad\frac{1}{f}\frac{d\phi}{dy}=-\partial_\phi
  W\,,\quad \frac{2(D-2)}{fg}\frac{dg}{dy}= W \,. \label{bog}
\end{align}
Note that one should not expect a Bogomol'nyi equation associated
to $f$ since it can be absorbed in a reparametrisation of the
transverse coordinate $y$.

The Bogomol'nyi equations can be solved by the domain wall
solution, generalising \cite{Cvetic:1999xx,Cvetic:2000zu}
 \begin{align}
   ds^2 & = h^{1/(2D-4)} d\vec{x}^2 + h^{(3-D)/(2D-4)} dy^2 \,, \notag \\
   M & = h^{1/n} \text{diag}(1/h_1,\ldots,1/h_n) \,, \qquad e^\phi = h^{-a/4}
   \,,
 \label{domain-wall}
 \end{align}
written in terms of the $n$ harmonic functions $h_i = 2 q_i y
+l_i^2$ and their product $h = h_1 \cdots h_n$.  Note that this
transverse coordinate basis has $\sqrt{-g} g^{tt} = -1$. The
functions $h_ i$ are necessarily positive since the entries of $M$
are postive. For $q_i>0$, this implies that $y$ can range from $0$
to $\infty$; if $q_i<0$, the range of $y$ is bounded from above.

The solution is parametrised by $n$ integration
constants\footnote{Strictly speaking, it is $l_i{}^2$ rather than
$l_i$ that appears as integration constant, allowing for positive
and negative $l_i{}^2$. However, one can always  take these
positive by shifting $y$, in which case the distinction between
$l_i$ and $l_i{}^2$ disappears \cite{Cvetic:1999xx}.} $l_i$.
However, if a charge $q_i$ happens to be vanishing, the
corresponding $l_i$ can always be set equal to one (by
$SL(n,\mathbb{R})$ transformations that leave the scalar potential
invariant). In addition, one can eliminate one of the remaining
$l_i$'s by a redefinition of the variable $y$. Therefore we
effectively end up with $p+q-1$ independent constants.

It should not be a surprise that all scalar potentials of
table~\ref{tab:gaugings}
satisfy the relation \eqref{Dna} since these are embedded in a
supergravity theory, whose Lagrangian ``is the sum of the
supersymmetry transformations'' and therefore always yields
first-order differential equations. For this reason, domain wall
solutions to the separate terms in \eqref{Bogomolnyi} will always
preserve half of supersymmetry. The corresponding Killing spinor
is given by
 \begin{align}
  \epsilon = h^{1/(8D-16)} \epsilon_0 \,, \qquad (1 + \Gamma_y) \epsilon_0 = 0 \,,
 \end{align}
where the projection constraint eliminates half of the components
of $\epsilon_0$. An exception is $a=0$, $q_i=1$ and $l_i = 0$, in
which case the domain wall solution \eqref{domain-wall} becomes a
maximally (super-)symmetric Anti-De Sitter space-time in
horospherical coordinates. In this case the singularity at $y=0$
is a coordinate artifact.

\section{Higher-dimensional Origin} \label{hd}

Upon uplifting the domain walls \eqref{domain-wall}, one obtains higher-dimensional
solutions, which are related to (near-horizon limits of) the 1/2 supersymmetric brane
solutions of 11D, IIA and IIB supergravity, as indicated in table~\ref{tab:gaugings}.
Note that the number of mass parameters (and therefore the number of harmonic functions
$h_i$ of the transverse coordinate) always equals the transverse dimension of the brane.
Thus, in $D$ dimensions, the number of mass parameters is related to the co-dimension of
the half-supersymmetric $(D-2)$-brane of IIA, IIB or M-theory.

The metric of the uplifted solution can in all cases be written in the form
 \begin{align}\label{brane}
  ds^2=H_n^{(2-n)/(D+n-3)}\,dx_{{\scriptscriptstyle{D-1}}}^2+H_n^{(D-1)/(D-n-3)}\,ds_{{\scriptscriptstyle{n}}}^2\,,
 \end{align}
where $H_n$ is a harmonic function on the transverse space, whose
powers are appropriate for the corresponding D-brane solution in
ten dimensions or M-brane solution in eleven dimensions. From the
form of the metric, it is therefore seen that the solution
corresponds to some kind of brane distribution. For all $q_i=1$,
these solutions were found in
\cite{Bakas:1999ax,Cvetic:1999xx,Bakas:1999fa} for the D3-, M2-
and M5-branes and in \cite{Cvetic:2000zu,Bakas:2000nt} for the
other non-conformal branes. The harmonic function takes the form
 \begin{align}\label{harm}
  H_n (y, \mu_i)=h^{-1/2} \Big( \sum_{i=1}^n\frac{q_i^2\mu_i^2}{h_i} \Big)^{-1} \,,
 \end{align}
where $\mu_i$ are Cartesian coordinates, fulfilling \eqref{hypersurface}. The
transverse part of the metric is given by \cite{Cvetic:1999xx}
 \begin{align}\label{transverse}
  ds_{{\scriptscriptstyle{n}}}^2 = H_n^{-1} h^{-1/2}dy^2+\sum_{i=1}^n h_i\,d\mu_i^2 \,.
 \end{align}
With a change of coordinates, it can be seen that the
$n$-dimensional transverse space is flat
\cite{Russo:1998mm,Cvetic:1999xx}
 \begin{align}\label{mutoz}
  z_i=\sqrt{h_i}\mu_i\,,\quad
  ds_{{\scriptscriptstyle{n}}}^2=\sum_{i=1}^n dz_i dz_i\,.
 \end{align}
The above is easily verified
\begin{align}
dz_i=h_i^{-1/2} q_i\mu_i\, dy+h_i^{1/2}\,d\mu_i\,,\quad
\sum_{i=1}^ndz_i dz_i=\sum_{i=1}^n
\frac{q_i^2\mu_i^2}{h_i}\,dy^2+\sum_{i=1}^n h_i\,d\mu_i^2\,,
\end{align}
where we have used $\sum_{i=1}^n q_i\,\mu_i\,d\mu_i=0$, which follows by from
\eqref{hypersurface}. Using \eqref{harm} it is seen that the above agrees with
\eqref{transverse}. Note that one has $\sqrt{-g} g^{ii}=1$ after the coordinate change to
$z_i$.

The harmonic function $H_n$ specifies the dependence on the $n$ transverse coordinates
$z_i$. The constants $l_i$ parametrise the possible harmonics that are consistent with
the reduction Ansatz. The mass parameters $q_i$ specify this reduction Ansatz. Thus,
changing a mass parameter $q_i$ changes both the reduction Ansatz and the harmonic
function that is compatible with that Ansatz. Sending a mass parameter to zero, e.g.~$q_n
\rightarrow 0$, corresponds to truncating the harmonic function on $n$-dimensional flat
space to
 \begin{align}
  H_n (q_n =0, l_n = 1) = H_{n-1} \,,
  \label{trunc-harm}
 \end{align}
i.e.~a harmonic function on $(n-1)$-dimensional flat space.

It is difficult to obtain the explicit expression for the harmonic
function $H_n$ in terms of the Cartesian coordinates $z_i$ (the
example of $n=2$ will be given in \eqref{D7-harmonic}).
Nevertheless, one can show that $H_n$ is indeed harmonic on
$\mathbb{R}^n$ for all values of $q_i$, thus extending the
analysis of \cite{Bakas:1999ax} where $q_i = 1$. The calculation
is facilitated by the following definitions
\begin{align}
A_m=\sum_{i=1}^n\frac{q_i^m z_i^2}{h_i^m}\,,\quad
B_m=\sum_{i=1}^n\frac{q_i^m}{h_i^m}\,.
\end{align}
In terms of $A_m$ and $B_m$ we calculate
\begin{align}
\partial_i H_n=h^{-1/2}\Big(-\frac{q_i z_i}{h_i}\frac{B_1}{A_2^2}+4\,\frac{q_i
z_i}{h_i}\frac{A_3}{A_2^3}-2\frac{q_i^2
z_i}{h_i^2}\frac{1}{A_2^2}\Big)\,,
\end{align}
from which we finally get
\begin{align}
\sum_{i=1}^n\partial_i\partial_i H_n=h^{-1/2}\Big(2 \frac{B_2}{A_2^2}-2\frac{B_1
A_3}{A_2^3}-16 \frac{A_4}{A_2^3}+16\frac{A_3^2}{A_2^4}-2
\frac{B_2}{A_2^2}+16\frac{A_4}{A_2^3}+2\frac{B_1
A_3}{A_2^3}-16\frac{A_3^2}{A_2^4}\Big)=0\,,
\end{align}
which proves the harmonicity of $H_n$ on $\mathbb{R}^n$.

The special case of the D6-brane solution \eqref{brane} with $n=3$ can be uplifted to one
dimension higher, where it becomes the 11D Kaluza-Klein monopole:
 \begin{align}
   ds^2 = dx_{7}^2 + H^{-1}(dy^3 + \sum_{i=1}^3 A_i dz_i)^2 + \sum_{i=1}^3 H dz_i{}^2 \,,
  \label{KK-monopole}
 \end{align}
where $y^3$ is the isometry direction of the KK-monopole. The functions $H=H(z_i)$ and
$A_i=A_i(z_j)$ are subject to the condition
 \begin{align}
  F_{ij} = \tfrac{1}{2}(\partial_i A_j - \partial_j A_i) =  \epsilon_{ijk} \partial_k H \,.
 \end{align}
Generally, this metric is the product of 7D Minkowski space-time and the 4D Euclidean
Taub-NUT space with isometry direction $y^3$.

The other special case is the D7-brane with $n=2$. Its harmonic function reads
 \begin{align}
  H_2 (y, \mu_i)= \Big( \sqrt{h_2} \mu_1{}^2 + \sqrt{h_1} \mu_2{}^2 \Big)^{-1} \,.
 \end{align}
In this case, it is straightforward (though perhaps not very insightful) to perform the
coordinate transformation to $z_i$, which yields:
 \begin{align}
  H_2(z_1,z_2) = \Big( \frac{\alpha z_1{}^2 + \beta z_2{}^2 + \gamma(z_1{}^2 + z_2{}^2)}{2 \gamma^2} \Big)^{1/2} \,,
 \label{D7-harmonic}
 \end{align}
with the definitions
 \begin{align}
  \begin{array}{cc}
  \alpha = q_1 q_2 (z_1{}^2 + z_2{}^2) + q_1 l_2^2 - q_2 l_1^2 \,, \\
  \beta = q_1 q_2 (z_1{}^2 + z_2{}^2) - q_1 l_2^2 + q_2 l_1^2 \,,
  \end{array} \qquad
  \gamma = \sqrt{\tfrac{1}{2}(\alpha^2 + \beta^2) + q_1 q_2 (z_1{}^2 -z_2{}^2)(\alpha-\beta)} \,.
 \end{align}
Indeed, it can be checked that this function is harmonic with respect to flat
$(z_1,z_2)$-space for all values of $q_i$ and $l_i$.

The special $SO(2)$-case with $q_1=q_2=1$ and $l_1=l_2$ leads to
the trivial harmonic function
\begin{equation}
H_2 = 1\, .
\end{equation}
Actually, for this case the 9D scalar potential is vanishing,
$V=0$, see also \eqref{V-single}. Although the scalar potential is
vanishing, the corresponding superpotential is non-vanishing: $W
\ne 0$. The corresponding 9D solution is uplifted to a 10D conical
spacetime. For more information, see \cite{Bergshoeff:2002mb}.

As another example, the $\mathbb{R}$-case with charges $(q_1,q_2) = (1,0)$ leads to
 \begin{align}
  H_2 (q_2 = 0) = |z_1| / l_2 \,,
\end{align}
which is a harmonic function in a one-dimensional transverse
space, in agreement with \eqref{trunc-harm}. This case describes
the ``circular'' D7-brane of \cite{Bergshoeff:1996ui} that is
T-dual to the D8-brane. Note that the previous case with $q_1=q_2$
has no manifest T-dual picture.

\section{Brane Distributions for $SO(n)$ Harmonics} \label{bd}

In this section we restrict ourselves to the $SO(n)$ cases and their group contractions.

Since the harmonic function $H_n$ depends on the angular variables
in addition to the radial, the uplifted solution will in general
correspond to a distribution of branes rather than a single brane.
For $D<9$ and $q_i=1$ (i.e.~the $SO(n)$ cases\footnote{Note that
the group contractions are included by taking some $q_i=0$, using
\eqref{trunc-harm}.} with $n \geq 3$) this means that the harmonic
function can be written in terms of a charge distribution $\sigma$
as follows \cite{Cvetic:1999xx,Cvetic:2000zu}
 \begin{align}\label{H-n}
  H_n(\vec{z}) = \int d^n z' \frac{\sigma(\vec{z}\,')}{\vert
  \vec{z}-\vec{z}\,'\vert^{n-2}}\ ,\ \qquad n\geq 3\,,
 \end{align}
and since $H_n$ appears without an integration constant, the distributions will actually
be a near-horizon limit of the brane distribution.

\begin{figure}[tb]
\centerline{\epsfig{file=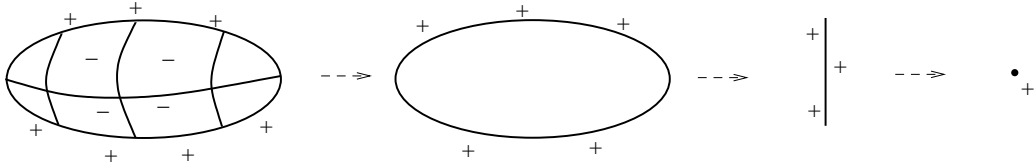,width=.85\textwidth}} \caption{\it The
distributions of D5-branes with three, two, one and zero non-vanishing $l_i$'s,
respectively.} \label{fig:D5-distrib}
\end{figure}

It turns out that the distributions are given in terms of higher dimensional ellipsoids
\cite{Kraus:1998hv,Cvetic:1999xx}. The dimension of these ellipsoids are given in
terms of the number $m$ of non-vanishing constants $l_i$. It is convenient
 to define
 \begin{align}
  x_m=1-\sum_{i=1}^m \frac{z_i^2}{l_i^2} \,, \qquad \vec{l} = ( l_1 , \ldots l_m , 0, \ldots , 0)
  \,,
 \end{align}
where the last $n-m$ constants $l_i$ are vanishing. Starting with
the case $m=n-1$, we have a negative charge\footnote{However,
these negative charges might be pathological, since the tension
will also be negative \cite{Freedman:1999gk,Dyson:2001kh}.}
distributed inside the ellipsoid and a positive charge distributed
on the boundary \cite{Freedman:1999gk,Cvetic:1999xx}:
 \begin{align}
  \sigma_{n-1} \sim \frac{1}{l_1 \cdots l_{n-1}}\, \Big(-x_{{\scriptscriptstyle{n-1}}}^{-3/2}
  \Theta(x_{{\scriptscriptstyle{n-1}}})+2\,
  x_{{\scriptscriptstyle{n-1}}}^{-1/2}
  \delta(x_{{\scriptscriptstyle{n-1}}})\Big)\,\delta^{(1)}(z_{{\scriptscriptstyle{n}}})
  \,.
  \label{double-distribution}
 \end{align}
Upon sending $l_{n-1}$ to zero, the charges in the interior of the
ellipsoid cancel, leaving one with a positive charge on the
boundary of a lower dimensional ellipsoid:
 \begin{align}
  \sigma_{n-2} \sim \frac{1}{l_1 \cdots l_{n-2}}\,
  \delta(x_{{\scriptscriptstyle{n-2}}})\,\delta^{(2)}(z_{{\scriptscriptstyle{n-1}}},z_{{\scriptscriptstyle{n}}})
  \,.
 \end{align}
Next, the contraction of more constants will yield brane
distributions over the inside of an ellipsoid. The distribution
$\sigma(z_i)$ is then a product of a delta-function and a
theta-function and the branes are localised along $n-m$
coordinates and distributed within an $m$-dimensional ellipsoid,
defined by $x_m=0$. For $1 \leq m \leq n-3$ non-zero constants,
one has
 \begin{align}
  \sigma_{m} \sim \frac{1}{l_1 \cdots l_{m}}\, x_ {{\scriptscriptstyle{m}}}^{(n-m-4)/2} \,
  \Theta(x_{{\scriptscriptstyle{m}}}) \,
  \delta^{(n-m)}(z_{{\scriptscriptstyle{m+1}}},\ldots,z_{{\scriptscriptstyle{n}}})
  \,.
 \end{align}
Finally, one is left with all constant $l_i$ vanishing, in which case the distribution
has collapsed to a point and generically reads
 \begin{align}
  \sigma_0 =
  \delta^{(n)}(z_{{\scriptscriptstyle{1}}},\ldots,z_{{\scriptscriptstyle{n}}})\,,
 \label{point-distribution}
 \end{align}
i.e.~we are left with (the near-horizon limit of)
a single brane. All these distributions satisfy
 \begin{align}
  \sigma_{m-1} = \delta(z_m) \int \sigma_m \,,
 \end{align}
consistent with the picture of distributions that collapse the $z_m$-coordinate upon
sending $l_m$ to zero. The case of D5-branes\footnote{For remarks about the corresponding
7D gauged supergravity, see the discussion.} is illustrated in figure
\ref{fig:D5-distrib}.

The uplift and the corresponding distributions were found in \cite{Cvetic:1999xx} for the
D3-, M2- and M5-branes. The extension to the other branes was treated in
\cite{Cvetic:2000zu}. Special cases have also been studied in \cite{Freedman:1999gk,
Dyson:2001kh}. For certain cases, i.e.~when the $l_i$'s are pairwise equal, the
distributions above are related to the extremal limit of rotating branes
\cite{Kraus:1998hv,Sfetsos:1998xd}, and the $l_i$'s then correspond to rotation
parameters.

%Q: what about the
%other parameters?

\begin{figure}[tb]
\centerline{\epsfig{file=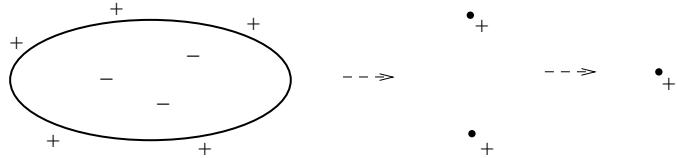,width=.55\textwidth}} \caption{\it The
distributions of D6-branes with  two, one and zero non-vanishing $l_i$'s, respectively.}
\label{fig:D6-distrib}
\end{figure}

The special case of the D6-brane distributions (with all $q_i=1$)
was first discussed in \cite{Cvetic:2000zu}. This splits up in
three separate possibilities, with $m=2, 1$ or $0$. The first
distribution $\sigma_2$ consists of positive and negative
densities and is given by the general formula
\eqref{double-distribution}. Upon sending $l_2$ to zero, this
collapses to
 \begin{align}
  \sigma_1 \sim \frac{1}{l_1} \,
  \delta \big( 1 - \frac{z_1{}^2}{l_1{}^2} \big) \, \delta^{(2)}(z_2 , z_3) \,.
 \end{align}
This is a distribution at the boundary of a one-dimensional ellipse, i.e.~it is localised at
the points $z_1 = \pm l_1$. For this reason, the corresponding harmonic function is given
by
 \begin{align}
  H_3 ({\vec z},l_1) = \frac{1}{2((z_1 - l_1)^2 + z_2{}^2 + z_3{}^2)^{1/2}} + \frac{1}{2((z_1 + l_1)^2 + z_2{}^2 +
  z_3{}^2)^{1/2}} \,,
 \end{align}
i.e.~the near-horizon limit of the double-centered D6-brane. Note that this is the only
non-trivial case (i.e.~with at least one $l_i$ non-vanishing) where we get a discrete
rather than a continuous distribution. Upon sending $l_1$ to zero, the brane distribution
$\sigma_0$ collapses to a point, as given in \eqref{point-distribution}. Indeed, the
harmonic function becomes
 \begin{align}
  H_3 = \frac{1}{|\vec{z}|} \,,
 \end{align}
i.e.~the near-horizon limit of the single-centered D6-brane with $SO(3)$-isometry. The
different distributions of D6-branes are shown in figure \ref{fig:D6-distrib}.

These explicit harmonic functions also give rise to special cases when uplifted to the
11D Kaluza-Klein monopole \eqref{KK-monopole}. The first consists of all $q_i =1$ and
only $l_1 \neq 0$, in which case the harmonic function corresponds to the near-horizon
limit of the two-centered solution. The Taub-NUT geometry degenerates to the
Eguchi-Hanson geometry in this case \cite{Eguchi:1979gw}. Secondly, the parameter $l_1$
can be sent to zero, yielding the near-horizon limit of a single-centered Kaluza-Klein
monopole. In this case, the 4D Euclidean space becomes (locally) flat.

In the other special case, with $n=2$ and $q_1 = q_2 = 1$, the IIB
solution can be understood as a distribution of D7-branes. Without
loss of generality we take $l_2 = 0$. Now \eqref{H-n} does not
apply and the harmonic function will instead be given by
 \begin{align}
  H_2({\vec z}) = 1 + \int dz_1' dz_2' \,\sigma_1(z_1', z_2'; l_1) \log((z_1-z_1')^2+(z_2-z_2')^2) \,,
  \label{D7-SO2-harmonic}
 \end{align}
with the D7-brane distribution
 \begin{align}
  \sigma_1 = \frac{1}{2 \pi l_1} \Big[
   - \big( 1- \frac{z_1'{}^2}{l_1 ^2} \big)^{-3/2} \Theta \big( 1- \frac{z_1'{}^2}{l_1^2} \big)
   + 2 \big( 1- \frac{z_1'{}^2}{l_1^2} \big)^{-1/2} \delta \big( 1- \frac{z_1'{}^2}{l_1^2} \big) \Big] \,.
 \label{D7-distribution}
 \end{align}
Note that this distribution consists of a line interval of
negative D7-brane density with positive contributions at both ends
of the interval. Both positive and negative contributions diverge
but these cancel exactly:
 \begin{align}
  \int dz_1' dz_2' \,\sigma_1(z_1', z_2') = 0 \,,
 \end{align}
i.e.~the total charge in the distribution \eqref{D7-distribution} vanishes.

\begin{figure}[tb]
\centerline{\epsfig{file=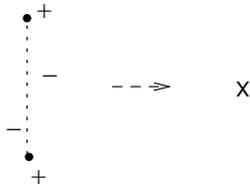,width=.2\textwidth}} \caption{\it The
distributions of D7-branes with one and zero non-vanishing $l_i$'s, respectively. The
cross indicates the conical singularity of the locally flat space-time.}
\label{fig:D7-distrib}
\end{figure}

The parameter $l_1$ of the general $SO(2)$ solution can be set to zero. This corresponds
to a collapse of the line interval to a point, as can be seen from
\eqref{D7-distribution}. However, due to the fact that the total charge vanishes, this
leaves us without any D7-brane density:
 \begin{align}
  \sigma_0 = 0 \,.
 \end{align}
This is consistent with the fact that the general harmonic
function \eqref{D7-harmonic} equals the one for the $SO(2)$ case
with $l_1 = l_2 = 0$. The D7-brane distributions are shown in
figure \ref{fig:D7-distrib}.

Thus, the two-dimensional $SO(2)$ harmonic function \eqref{D7-SO2-harmonic} of the
D7-brane differs in two important ways from the generic $SO(n)$ harmonic function with
$n>2$. Firstly, the total charge distribution of D7-branes vanishes, while it adds up to
a finite and positive number in the other cases. Secondly, but not unrelated, one needs
to include a constant in the harmonic function \eqref{D7-SO2-harmonic} in terms of the
distribution. In the generic cases this constant was absent, corresponding to the
near-horizon limit of these branes. For the D7-brane, the concept of a near-horizon limit
is unclear, as we comment upon in the discussion.

\section{Discussion} \label{discussion}

In this paper we have discussed the construction of $CSO$ gauged supergravities via
dimensional reduction over a coset manifold, a group manifold or with a twist. In
addition, half-supersymmetric domain wall solutions were constructed. The uplifting of
these solutions to string and M-theory leads to brane solutions characterized by a
harmonic function $H_n$ on the flat transverse space $\mathbb{R}^n$. At this point our
results apply for arbitrary values of the charges $q_i$ thereby generalizing the case of
all $q_i=1$ \cite{Cvetic:1999xx, Cvetic:2000zu}. The rewriting of the harmonic function
$H_n$ in terms of an integral representation with a brane distribution function
$\sigma_m({\vec z})$ was discussed for the $SO(n)$ case and its group contractions,
i.e.~all $q_i \ge 0$, where special emphasis was put on the $SO(2)$ and $SO(3)$ cases. It
would be interesting, if possible, to extend this part of the discussion to arbitrary
values of the charges.

We would like to comment on the special features of the $SO(2)$ harmonic function
corresponding to the D7-brane that we encountered in section~\ref{bd}. The near-horizon
limit of D-branes does not yield a separated spherical part in Einstein frame; for this
one needs to go to the so-called dual frame, in which the tension of the brane is
independent of the dilaton
 \begin{align}
  g_{\mu \nu}^{\text{dual}} = \exp \left( \frac{(3-p)}{2(p-7)} \, \phi  \right) g_{\mu \nu}^{\text{Einstein}}
  \,.
 \end{align}
In the dual frame, the near-horizon geometry of all D-branes with
$p \leq 6$ reads\footnote{Except for the case $p = 5$, which has
Minkowski$_7$ rather than $AdS_7$ \cite{Gibbons:1993sv}.}
$AdS_{p+2} \times S^{n-1}$. Clearly, this formula does not hold
for the D7-brane; a related complication is the fact that the dual
object is the D-instanton, which lives on a Euclidean space.

One of the themes of this paper has been that for every brane there is a corresponding
gauged supergravity. Sofar, however, we have not treated all the branes of string and
M-theory. For one thing, we did not explicitly include the NS5A-brane in
table~\ref{tab:gaugings}. Since this brane is the direct dimensional reduction of the
M5-brane, it leads to the $D=7$ $ISO(4)$ gauged supergravity of \cite{Cvetic:2000ah}.
Only the $SO(4)$ symmetries are linearly realized, the other four symmetries occur as
Stueckelberg symmetries. A similar story holds for the D2-brane which is the direct
dimensional reduction of the M2-brane. Furthermore, we did not consider the IIB doublets
of NS5B/D5- and F1B/D1-branes. The associated theories are the reduction of IIB over
$S^3$ or $S^7$ with an electric or magnetic flux of the NS-NS/R-R three form
field-strength \cite{Cvetic:2000dm, Cvetic:2003jy}. For the $D=3$ $SO(8)$ theories
corresponding to the IIB strings (which are different than the F1A result), see
\cite{Fischbacher:2003yw}. The five-brane cases are supposed to lead to new $D=7$ $SO(4)$
gauged supergravities, which might be related to the theories constructed in
\cite{Alonso-Alberca:2002tb}.

Finally, we would like to mention that there are also gauged maximal supergravities which
are not of the $CSO$ class considered in this paper, see e.g.~\cite{Andrianopoli:2002mf,
Hull:2002cv, deWit:2002vt}. In addition, there are gauged maximal supergravities that do
not have an action but only have field equations \cite{Bergshoeff:2002nv,
Bergshoeff:2003ri}; such theories do not allow for domain wall solutions, however. It
would be interesting to investigate the structure of such gaugings and to establish
whether the absence of domain wall solutions is a general feature of these theories.

\section*{Acknowledgements}

We thank Hermann Nicolai and Tom\'{a}s Ort\'{\i}n for interesting discussions. This work
is supported in part by the Spanish grant BFM2003-01090 and by the European Community's
Human Potential Programme under contract HPRN-CT-2000-00131 Quantum Spacetime, in which
E.B., M.N.~and D.R.~are associated to Utrecht University.

\appendix

\section{Group Contraction and Dimensional Reduction}\label{contraction}

We would like to consider two operations on the scalar sector of
the $CSO$ gauged supergravity. The first operation corresponds to
a contraction of the $CSO$ gauge group and corresponds to setting
one mass parameter equal to zero. For concreteness it is taken to
be the last one: $q_i = (q_p , 0)$, where we have split up
$i=(p,n)$ and $p = 1,\ldots,n-1$. The superpotential now reads
 \begin{align}
  W = e^{a \phi/2} \sum_{p} q_p\, e^{\vec{\alpha}_{p} \cdot \vec{\phi}}
   = e^{a \phi/2 + \vec{\beta} \cdot \vec{\phi}} \sum_p q_p\, e^{\vec{\beta}_p \cdot \vec{\phi}}
   \,,
  \label{contracted-superpotential}
 \end{align}
where we have chosen to split off an overall part $\vec{\beta} \cdot \vec{\phi}$
according to $\vec{\alpha}_p = \vec{\beta} + \vec{\beta}_p$. A convenient choice for
$\vec{\beta}$ is
 \begin{align}
  \vec{\beta} = - \frac{1}{n-1} \vec{\alpha}_n = (0,\ldots,0,\frac{1}{\sqrt{n(n-1)/2}})
  \,.
 \end{align}
This corresponds to the scalar coset split
 \begin{align}
  M = \left(%
 \begin{array}{cc}
  e^{\vec{\beta} \cdot \vec{\phi}} \tilde{M} & 0 \\
  0 & e^{-(n-1) \vec{\beta} \cdot \vec{\phi}} \\
 \end{array}%
 \right) \,, \qquad
  \tilde{M} = \text{diag}(e^{\vec{\beta}_1 \cdot \vec{\phi}} , \ldots , e^{\vec{\beta}_{n-1} \cdot
  \vec{\phi}})\,,
 \end{align}
where the weight vectors $\vec{\beta}_p$ are subject to the
reduction of \eqref{weights}:
 \begin{align}
  \sum_p \beta_{pI}=0 \,, \qquad
  \sum_p \beta_{pI}\,\beta_{pJ} = 2 \, \delta_{IJ} \,, \qquad
  \vec{\beta}_{p} \cdot \vec{\beta}_{q} = 2 \, \delta_{pq}-\frac{2}{n-1} \,,
  \label{weights-beta}
 \end{align}
while the last component of all vectors $\vec{\beta}_p$ vanishes:
$\beta_{pn} = 0$. Therefore, the contracted superpotential
\eqref{contracted-superpotential} only depends on the smaller
coset $SL(n-1,\mathbb{R}) / SO(n-1)$. Also note that the overall
dilaton coupling has changed due to the contraction. For the
scalar potential, this will amount to $a \phi + 2 \vec{\beta}
\cdot \vec{\phi}$ instead of $a \phi$. After a change of basis,
corresponding to an $SO(n+1)$ rotation in
$(\phi,\vec{\phi})$-space, this takes the form $\tilde{a}
\tilde{\phi}$ with
 \begin{align}
  \tilde{a}^2 = a^2 + 4 \vec{\beta} \cdot \vec{\beta} = \frac{8}{n-1} - 2 \, \frac{D-3}{D-2} > a \,,
  \label{Dna-contracted}
 \end{align}
which is exactly the original relation \eqref{Dna} with $n$ decreased by one. It should
be clear that this contraction can be employed several times, each time reducing $n$ by
one.

The second operation we wish to perform corresponds to
dimensionally reducing the scalar sector. We take trivial
Ans\"atze\footnote{In the reduction Ans\"atze, hatted quantities
are $D$-dimensional, while unhatted ones are $(D-1)$-dimensional.}
for the scalars, $\hat{M} = M$ and $\hat{\phi} = \phi$, and the
usual Ansatz for the metric (obtaining Einstein frame with a
canonically normalised Kaluza-Klein scalar $\varphi$ in the lower
dimension):
 \begin{align}
  \hat{ds}{}^2 = e^{2 \gamma \varphi} ds^2 + e^{- 2 (D-3) \gamma \varphi} dz^2 \,, \qquad
  \gamma^2 = \frac{1}{2(D-2)(D-3)} \,,
 \label{Ansatz-gravity-circle}
 \end{align}
where we have truncated the Kaluza-Klein vector away. The
resulting scalar potential is of the same form
\eqref{CSO-potential}, but again the dilaton coupling has changed:
the factor $a \phi$ is replaced by $a \phi + 2\gamma \varphi$.
After a field redefinition, this corresponds to $\tilde{a}
\tilde{\phi}$ with
 \begin{align}
  \tilde{a}^2 = a^2 + 4 \gamma^2 = \frac{8}{n} - 2 \, \frac{D-4}{D-3} > a \,,
  \label{Dna-reduced}
 \end{align}
which is exactly the original relation \eqref{Dna} with $D$ decreased by one. Again,
dimensional reduction can be performed any number of times, reducing $D$ by one at each
step.

\bibliography{dw}
\bibliographystyle{utphysmodb}

\end{document}